\documentclass[aps,prl,reprint,amsmath,amssymb,superscriptaddress,floatfix,longbibliography]{revtex4-1}
\usepackage{amsthm,dsfont,ctable,url,braket,mathtools}
\usepackage{natbib}
\usepackage[colorlinks,
            linkcolor=blue,
            anchorcolor=blue,
            citecolor=blue,
urlcolor=blue
            ]{hyperref}
\usepackage{graphicx}
\usepackage{float}

\usepackage[justification=raggedright]{caption}
\usepackage{subcaption}

\usepackage{ulem}
\usepackage[capitalise]{cleveref}

\usepackage[T1]{fontenc}
\usepackage{lmodern}

\newcommand{\Tr}{\operatorname{Tr}}

\definecolor{ag}{rgb}{0.0, 0.5, 0.0}

\newcommand{\RNum}[1]{\uppercase\expandafter{\romannumeral #1\relax}}

\newcommand{\tr}{\operatorname{Tr}}
\newcommand{\val}[1]{\langle#1\rangle}
\newcommand{\lb}{\left\{}
\newcommand{\rb}{\right\}}

\newcommand{\nc}{\newcommand}
 \nc{\bA}{{\bf A}} \nc{\bB}{{\bf B}} \nc{\bC}{{\bf C}}
 \nc{\bD}{{\bf D}} \nc{\bE}{{\bf E}} \nc{\bF}{{\bf F}}
 \nc{\bG}{{\bf G}} \nc{\bH}{{\bf H}} \nc{\bI}{{\bf I}}
 \nc{\bJ}{{\bf J}} \nc{\bK}{{\bf K}} \nc{\bL}{{\bf L}}
 \nc{\bM}{{\bf M}} \nc{\bN}{{\bf N}} \nc{\bO}{{\bf O}}
 \nc{\bP}{{\bf P}} \nc{\bQ}{{\bf Q}} \nc{\bR}{{\bf R}}
 \nc{\bS}{{\bf S}} \nc{\bT}{{\bf T}} \nc{\bU}{{\bf U}}
 \nc{\bV}{{\bf V}} \nc{\bW}{{\bf W}} \nc{\bX}{{\bf X}}
 \nc{\bZ}{{\bf Z}}

\nc{\cA}{{\cal A}} \nc{\cB}{{\cal B}} \nc{\cC}{{\cal C}}
\nc{\cD}{{\cal D}} \nc{\cE}{{\cal E}} \nc{\cF}{{\cal F}}
\nc{\cG}{{\cal G}} \nc{\cH}{{\cal H}} \nc{\cI}{{\cal I}}
\nc{\cJ}{{\cal J}} \nc{\cK}{{\cal K}} \nc{\cL}{{\cal L}}
\nc{\cM}{{\cal M}} \nc{\cN}{{\cal N}} \nc{\cO}{{\cal O}}
\nc{\cP}{{\cal P}} \nc{\cQ}{{\cal Q}} \nc{\cR}{{\cal R}}
\nc{\cS}{{\cal S}} \nc{\cT}{{\cal T}} \nc{\cU}{{\cal U}}
\nc{\cV}{{\cal V}} \nc{\cW}{{\cal W}} \nc{\cX}{{\cal X}}
\nc{\cZ}{{\cal Z}}
\def\){ \right)}
\def\({\left(}

\begin{document}

\title{Observing geometry of quantum states in a three-level system}

\author{Jie Xie}
\altaffiliation{These authors contributed equally to this work.}
\affiliation{National Laboratory of Solid State Microstructures, Key Laboratory of Intelligent Optical Sensing and Manipulation (Ministry of Education), College of Engineering and Applied Sciences and School of Physics, Nanjing University, Nanjing 210093, China}
\affiliation{Collaborative Innovation Center of Advanced Microstructures, Nanjing University, Nanjing 210093, China}
\author{Aonan Zhang}
\altaffiliation{These authors contributed equally to this work.}
\affiliation{National Laboratory of Solid State Microstructures, Key Laboratory of Intelligent Optical Sensing and Manipulation (Ministry of Education), College of Engineering and Applied Sciences and School of Physics, Nanjing University, Nanjing 210093, China}
\affiliation{Collaborative Innovation Center of Advanced Microstructures, Nanjing University, Nanjing 210093, China}

\author{Ningping Cao}
\affiliation{Department of Mathematics $\&$ Statistics, University of Guelph, Guelph N1G 2W1, Ontario, Canada}
\affiliation{Institute for Quantum Computing, University of Waterloo, Waterloo N2L 3G1, Ontario, Canada}

\author{Huichao Xu}
\affiliation{National Laboratory of Solid State Microstructures, Key Laboratory of Intelligent Optical Sensing and Manipulation (Ministry of Education), College of Engineering and Applied Sciences and School of Physics, Nanjing University, Nanjing 210093, China}
\affiliation{Collaborative Innovation Center of Advanced Microstructures, Nanjing University, Nanjing 210093, China}

\author{Kaimin Zheng}
\affiliation{National Laboratory of Solid State Microstructures, Key Laboratory of Intelligent Optical Sensing and Manipulation (Ministry of Education), College of Engineering and Applied Sciences and School of Physics, Nanjing University, Nanjing 210093, China}
\affiliation{Collaborative Innovation Center of Advanced Microstructures, Nanjing University, Nanjing 210093, China}

\author{Yiu-Tung Poon}
\affiliation{Department of Mathematics, Iowa State University, Ames, Iowa, IA 50011, USA}

\author{Nung-Sing Sze}
\affiliation{Department of Applied Mathematics, The Hong Kong Polytechnic University, 999077 Hong Kong, China}

\author{Ping Xu}
\affiliation{National Laboratory of Solid State Microstructures, Key Laboratory of Intelligent Optical Sensing and Manipulation (Ministry of Education), College of Engineering and Applied Sciences and School of Physics, Nanjing University, Nanjing 210093, China}
\affiliation{Collaborative Innovation Center of Advanced Microstructures, Nanjing University, Nanjing 210093, China}
\affiliation{Institute for Quantum Information $\&$ State Key Laboratory of High Performance Computing, College of Computer, National University of Defense Technology, Changsha 410073, China}

\author{Bei Zeng}
\email[]{zengb@ust.hk}
\affiliation{Department of Physics, The Hong Kong University of Science and Technology, Clear Water Bay, Kowloon, 999077 Hong Kong, China}

\author{Lijian Zhang}
\email[]{lijian.zhang@nju.edu.cn}
\affiliation{National Laboratory of Solid State Microstructures, Key Laboratory of Intelligent Optical Sensing and Manipulation (Ministry of Education), College of Engineering and Applied Sciences and School of Physics, Nanjing University, Nanjing 210093, China}
\affiliation{Collaborative Innovation Center of Advanced Microstructures, Nanjing University, Nanjing 210093, China}

\date{\today}

\begin{abstract}
In quantum mechanics, geometry has been demonstrated as a useful tool for inferring non-classical behaviors and exotic properties of quantum systems. One standard approach to illustrate the geometry of quantum systems is to project the quantum state space to the Euclidean space via measurements of observables on the system. Despite the great success of this method in studying two-level quantum systems (qubits) with the celebrated Bloch sphere representation, there is always the difficulty to reveal the geometry of multi-dimensional quantum systems. Here we report the first experiment measuring the geometry of such projections beyond the qubit. Specifically, we observe the joint numerical ranges (JNRs) of a triple of observables in a three-level photonic system, providing complete classification of the JNRs. We further show that the geometry of different classes reveal ground-state degeneracies of a Hamiltonian as a linear combination of the observables, which is related to quantum phases in the thermodynamic limit. Our results offer a versatile geometric approach for exploring the properties of higher-dimensional quantum systems.
\end{abstract}
\maketitle

\par
Arising from Euclid's first attempt of establishing its axiomatic form, geometry has become an essential method for understanding the physical world, especially in the field of quantum physics~\cite{bengtsson_zyczkowski_2006,PhysRevLett.65.1697,Nielsen1133,Flaschner2016}. An exemplary use of geometric method is in the creation of statistical mechanics in 1870s, when Gibbs introduced a geometric means to infer thermodynamic properties (e.g. energy or entropy) of a system by considering a convex body constituted by all possible values of physical quantities~\cite{gibbs1873method}. Shifting to systems that the behaviors are governed by quantum physics, the possible expectation values of physical quantities are instead acquired over the entire space of quantum states, mathematically the set of all semi-definite matrices of trace one $\mathcal{M}_d=\{\rho:\rho \geq 0,\ \Tr(\rho) = 1\}$ in a $d$-dimensional Hilbert space. The convex body formed by joint expectation values of different observables on all quantum states can be used as a geometric representation of quantum state space. One of the most successful example is the Bloch sphere of qubit state space \cite{Kimura2003}, which has become the fundamental model in quantum information~\cite{nielsen2001quantum}. Many works are devoted to studying the geometry of higher-dimensional systems, such as generalizing the Bloch vector \cite{Kimura2003,Arvind_1997,Bertlmann_2008,aerts2014extended} and visualizing single qutrit state~\cite{PhysRevA.93.062126}. However, there is no satisfactory geometric ways to visualize the whole higher-dimensional state space, which possess properties dissimilar to those of qubits and start to play indispensable roles in quantum information processing~\cite{lanyon2009simplifying,zhang2016engineering,wang2018multidimensional}.
\par
In this work, we investigate the geometry of quantum states by projecting the state space $\mathcal{M}_d$ onto $n$-dimensional Euclidean space $\mathbb{R}^n$ via measurements of $n$ observables on the system~\cite{Toeplitz1918,Hausdorff1919,Dunkl_2011,Gutkin2013}. This geometric construction allows exploration of the complicated structure and physical properties of high-dimensional quantum systems through their lower-dimensional projections. Behind this construction is the concept of numerical range in mathematical terminology. Back in 1918, Toeplitz~\cite{Toeplitz1918} introduced the numerical range of a $d\times d$ complex matrix $F$, which is defined as $W(F)=\{z=\langle \psi |F|\psi\rangle:|\psi\rangle \in \mathbb{C}^d, \langle \psi|\psi\rangle=1\}$. Here $F=F_1+ \mathrm{i} F_2$ involves two Hermitian matrices $F_1$ and $F_2$. The conjecture by Toeplitz that $W(F)$ is convex was later proved by Hausdorff in 1919~\cite{Toeplitz1918,Hausdorff1919}.
A natural extension is termed as joint numerical range (JNR)~\cite{Gutkin2013} involving a collection of Hermitian matrices $\mathcal{F}=\{F_1,...,F_n\}$,
\begin{equation}
W(\mathcal{F})=\{(\langle \psi |F_1|\psi\rangle,...,\langle \psi |F_n|\psi\rangle):|\psi\rangle \in \mathbb{C}^d, \langle \psi|\psi\rangle=1\},
\end{equation}
naturally forming a geometric object in $\mathbb{R}^n$. Then the state space projection via these matrices, which allows statistical mixture of pure states $|\psi\rangle$, is simply the convex hull of the JNR
\begin{equation}
L(\mathcal{F})=\{(\Tr(\rho F_1),...,\Tr(\rho F_n)):\rho \in \mathcal{M}_d\}.
\end{equation}
In the following, we are mainly concerned with the set $L(\mathcal{F})$ and do not distinguish it from $W(\mathcal{F})$ (see Supplemental Material Sec. \RNum{1} A \cite{Supplementary}). 
\par
In recent years, the topic of numerical range has been reviewed in the study of quantum phase transition~\cite{Chen_2015,Zauner_2016,PhysRevA.93.012309,Spitkovsky2018}, deriving uncertainty relations \cite{Szyma_ski_2019} and entanglement criterion \cite{PhysRevA.100.042326} and error correcting codes in quantum computing~\cite{10.1080/03081080802677441,li2011generalized}. Yet many characteristics of JNR are still unknown for dimension as low as 3. Recently, Szyma{\'n}ski, Weis and {\.Z}yczkowski made a crucial step towards this problem by proving complete classification of the JNR in the case $d=n=3$~\cite{szymanski2018classification}. However, experimentally recovering the geometry of JNR, together with its classification, demands sampling adequate data, which is a non-trivial problem requiring the ability to implement arbitrary unitary operations on a qudit system~\cite{PhysRevLett.73.58,Carolan711}. Here we perform the first experiment allowing complete observation of state space projections beyond qubit systems.
\begin{figure}
\centering
\includegraphics[width=\linewidth]{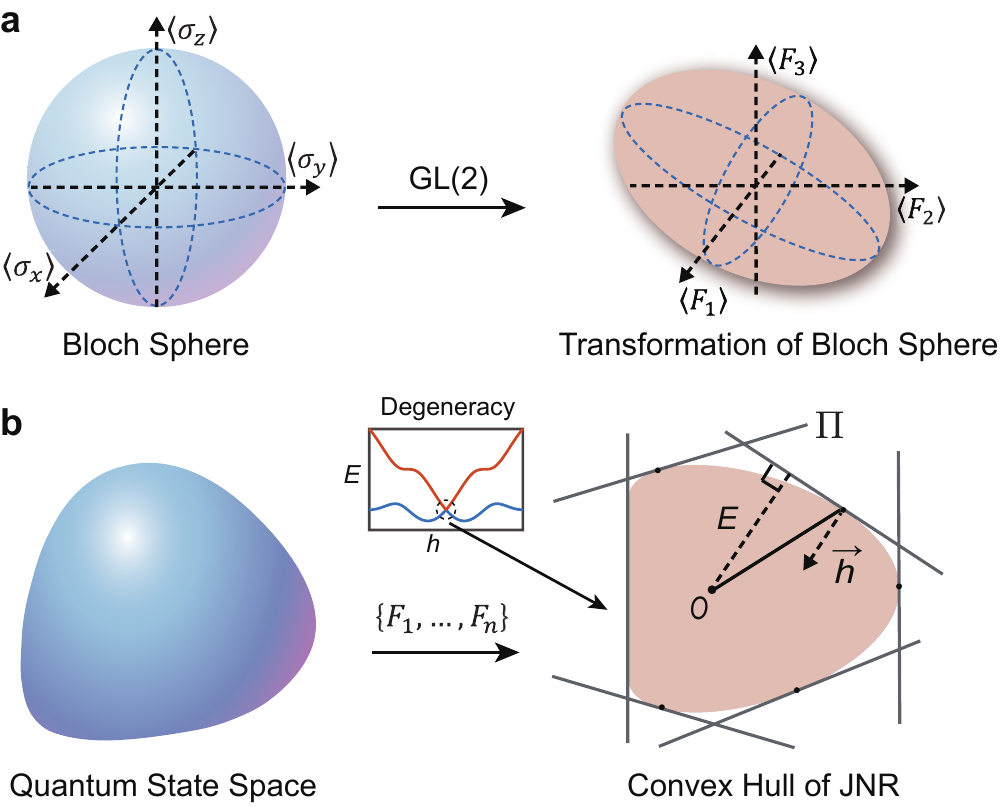}
\caption{\label{fig:intro}Quantum state space projection and the joint numerical range (JNR). (a) The state space of a qubit system $\mathcal{M}_2$ can be represented as a Bloch sphere, which is the JNR of the three Pauli operators $\{\sigma_x,\sigma_y,\sigma_z\}$. Undergoing a linear transformation $GL(2)$, the Bloch sphere becomes an ellipsoid, as the JNR of other three linearly independent matrices. (b) The state space of a higher-dimensional quantum system ($d\geq3$) is a convex set with a more complicated structure. Following the qubit case, these structures can be revealed by projecting the state space $\mathcal{M}_d$ to $\mathbb{R}^n$ through a set of $n$ Hermitian observables. This constitute the convex hull of the JNR, whose surface can be generated by the ground-states of a set of system Hamiltonians $H(\vec{h})$, geometrically corresponding to a set of supporting hyperplanes $\Pi$ (grey, solid lines). In general, ground-state degeneracy happens when a flat portion appears on the surface.}
\end{figure}

\par
As mentioned above, a simple example in qubit system is the JNR of three Pauli operators known as the Bloch sphere~\cite{Kimura2003}, whereas the JNR of other Hermitian matrices is equivalent to a transformation of the Bloch sphere, as shown in Fig. \ref{fig:intro}(a). Extended to higher-dimensional systems, the geometry of JNR is associated with a class of system Hamiltonians $H(\vec{h})=\sum h_i F_i$ parameterized by $\vec{h}=(h_1,...,h_n)$ in the basis $\mathcal{F}$. It is intuitive that surface points (extreme points) of JNR are genetrated by ground-states of $H(\vec{h})$ (see Supplemental Material Sec. \RNum{1} B \cite{Supplementary}).
Geometrically, each parameter vector $\vec{h}$ represents a inward-pointing unit vector in $\mathbb{R}^n$ and corresponds a supporting hyperplane $\Pi$ tangent to the surface of JNR, as depicted in Fig. \ref{fig:intro}(b). The tangent point is acquired by the ground-state $\ket{\psi_g(\vec{h})}$ of $H(\vec{h})$, while the corresponding ground-state energy $E_g$ can be obtained by projecting the point $(\langle F_1\rangle_{\psi_g(\vec{h})},...,\langle F_n\rangle_{\psi_g(\vec{h})})$ onto the $\vec{h}$ direction. When continuously varying the parameter vector $\vec{h}$, the envelope of all supporting hyperplane constitutes the surface of JNR, which can be generated by all ground-states $\ket{\psi_g(\vec{h})}$. If there exists a flat portion on the surface of JNR at a particular direction $\vec{h}$, the supporting hyperplane will be tangent to the whole portion instead of a single point. Therefore, this indicates ground-state degeneracy in the sense that different ground-states are associated with one system Hamiltonian $H(\vec{h})$~\cite{Spitkovsky2018}.
\par
In the case $d=n=3$, the image of the JNR forms a three-dimensional oval that can be classified in terms of its one- or two-dimensional faces, that is, segments or filled ellipses. These faces are invariant under linear transformation and translation. According to the number of segments ($s$) and filled ellipses ($e$) on the surface, all the JNRs can be divided into ten possible categories \cite{szymanski2018classification}, among which the eight unitarily irreducible categories that we measured are as follows (the other two categories correspond to linearly dependent operator sets, which can be derived by lower-dimensional JNRs, see Supplemental Material Sec. \RNum{2} D \cite{Supplementary}):
\begin{equation*}
s=0;e=0,1,2,3,4 \text{ and } s=1;e=0,1,2. 
\end{equation*}

\begin{figure*}
\centering
\includegraphics[width=\linewidth]{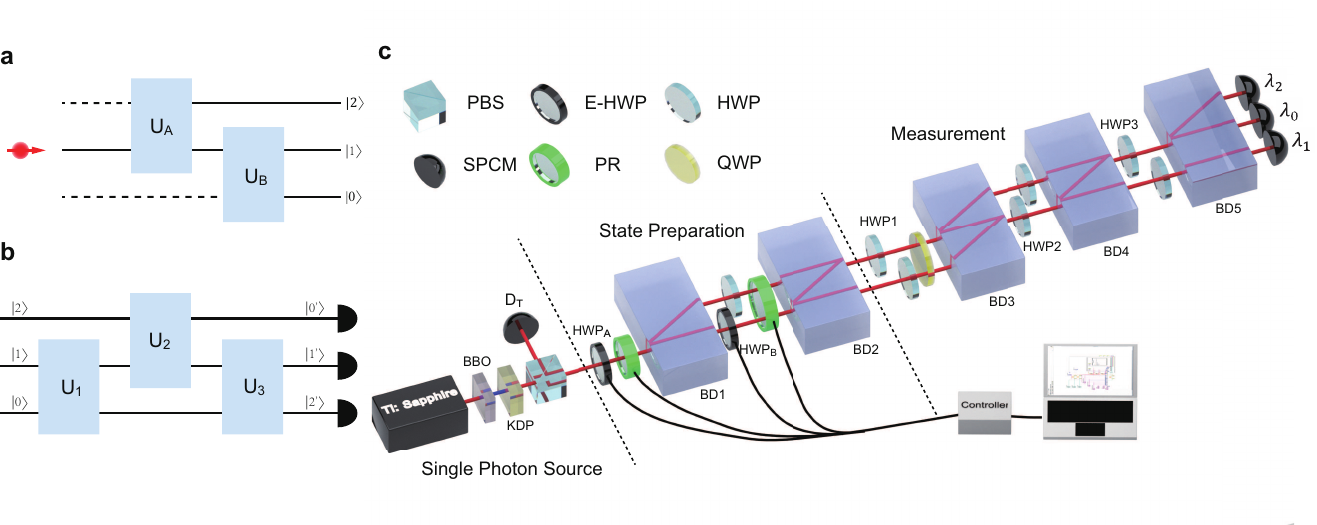}
\caption{\label{fig:setup}Experimental scheme. (a) Preparation of single photonic qutrit state. A single photon propagating in three modes represents a three-level quantum system. Sequential two-mode unitary operations can evolve the system to an arbitrary superposition of the three levels. (b) Measurement of a Hermitian observable $F_i$. The three-outcome measurement is constituted by three two-modes unitary transformations followed by single-photon detections, which projects input states onto the three eigenstates of the observable $F_i$. (c) Experimental set-up. Photon pairs are generated through the parametric downconversion process in a phase-matched potassium di-hydrogen phosphate (KDP) crystal pumped by frequency-doubled Ti:Sapphire laser pulses and then separated by a polarizing beam splitter (PBS). A single photon in the transmitted path is heralded by detection of a reflected photon at the heralding detector $D_{T}$. The state preparation module is composed of two electronically-controlled half-wave plates (E-HWPs), two phase retarders (PRs) and two calcite beam displacers (BDs). The measurement part is composed of wave plates, BDs and three single photon counting modules (SPCMs). For some observables, a quarter-wave plate (QWP) is inserted before BD3. Unlabelled HWPs are set to $45^\circ$ or $0^\circ$.}
\end{figure*}

\par
To measure the JNR, one can prepare identical copies of quantum states and obtain a triple of expectation values of $\mathcal{F}$ by separately measuring each of the three observables on the same states. Experimentally, qutrit states can be encoded in photons' different degrees of freedom \cite{PhysRevLett.91.227902,PhysRevA.70.042303,PhysRevA.83.051801,lapkiewicz2011experimental}. Here we prepare a single photon in the superposition of three modes which is equivalent to a three-level system, such as a spin-one particle. The single photon, generated in a heralded manner by the parametric downconversion process, is initially injected into one of the three modes. Then after two sequential two-mode unitary operations (Fig. \ref{fig:setup}(a)), the system is prepared in any pure superposition of the three levels, of the form $\ket{\psi}=\cos\theta_1e^{i\phi_1}\ket{0}+\sin\theta_1\sin\theta_2e^{i\phi_2}\ket{1}+\sin\theta_1\cos\theta_2\ket{2}$. This operation is realized by the state preparation module in Fig. \ref{fig:setup}(c), which involves wave-plates together with a calcite beam displacer (BD) to distribute the single photons among three optical modes and two variable phase retarders to manipulate phases between different modes. The three optical modes are defined as the horizontal and vertical polarizations of a single path mode (top) and the vertical polarization (the horizon polarization is not used) of another path mode (lower). The measurement of any Hermitian observable (Fig. \ref{fig:setup}(b)) is realized by a three-outcome quantum measurement collapsing the state onto one of the three eigenstates of the observable $F_i$. The state firstly undergoes a unitary evolution consisting of three sequential two-mode unitaries, so that the input state is transformed in the eigen-basis of the observable $F_i$~\cite{PhysRevLett.73.58}. This transformation is accomplished by three cascaded interferometers formed by a set of wave-plates and BDs, as shown in the measurement part of Fig. \ref{fig:setup}(c). At the end three single photon counting modules (SPCMs) are used for the detection of single photons. Clicks of the three detectors indicate measuring the corresponding eigenvalues $\lambda_j^{(i)}(j=0,1,2)$ of $F_i$, so that the expectation values of $F_i$ can be estimated through the measured probability distribution $\{p_j\}$. The wave-plates in the measurement stage can be configured in different settings, thus enabling measurements of different observables with respect to the same set of states.
\begin{figure*}
\centering
\includegraphics[width=\linewidth]{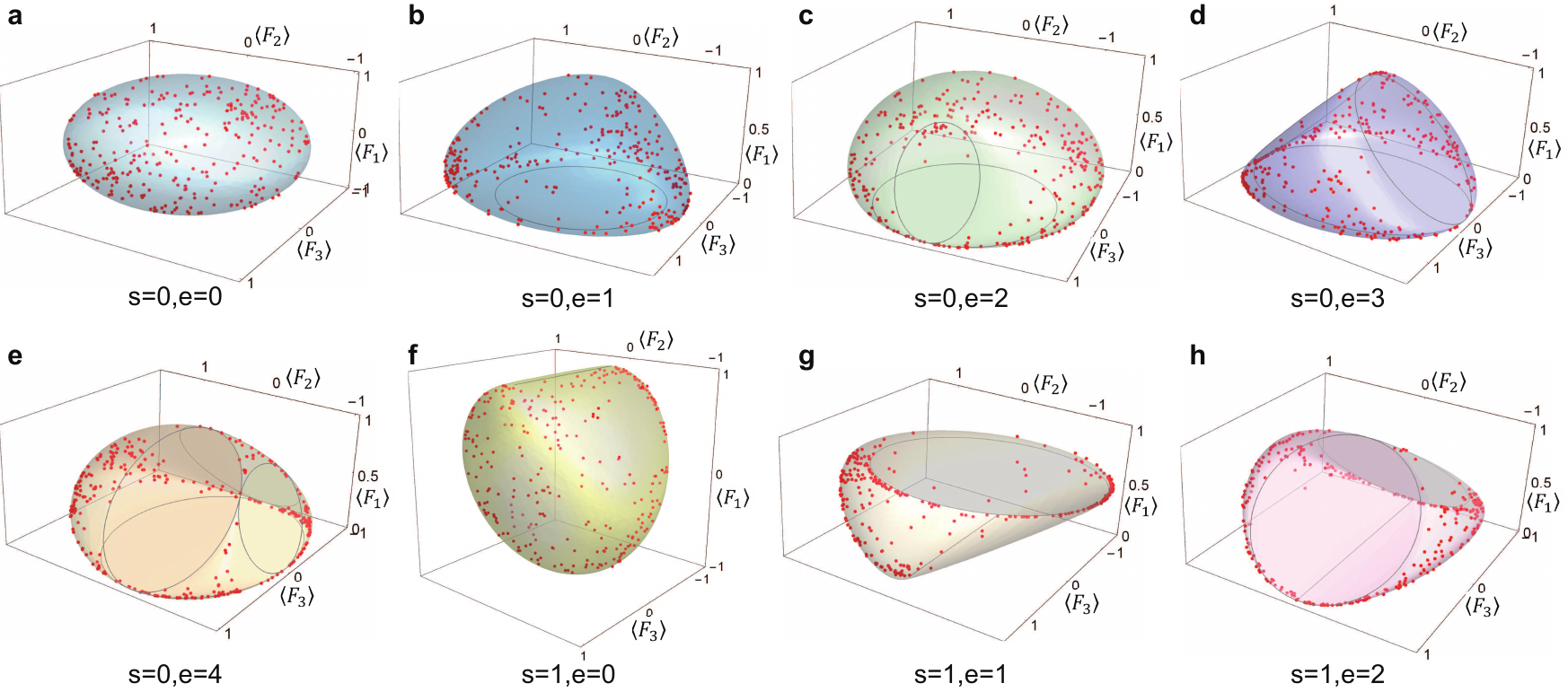}
\caption{\label{fig:exp_results}Complete observation of the joint numerical range (JNR). (a)-(h) Experimentally measured data (red dots) and the theoretical predictions of JNR (chromatic convex bodies) for the examples of the eight possible classes. Each class is specified by the number of segments ($s$) and filled ellipses ($e$) on the surface (ploted with gray lines) of the convex body. The experimental data are obtained by measuring the surface states (ground-states) of the system Hamiltonians (Eq. (\ref{eqn:Hamiltonian})). For each class, 300 surface states are sampled. The theoretical bodies are plotted by sampling adequate ground-states (more than 3000 for each class) and then generating the convex hull of the corresponding theoretical points of the JNR.}
\end{figure*}
\par
For each class of JNR we provide an example of the 3-tuple of observables $\mathcal{F}$ (see Supplemental Material Sec. \RNum{2} B \cite{Supplementary}) being measured in our experiment. Given the set of JNR is convex for $n=3$~\cite{Au-Yeung197985} and any interior point can be obtained by the mixture of surface states, measuring pure surface states is enough for the observation of JNR. We randomly sample 300 ground-states $\ket{\psi_g(\theta,\phi)}$ of system Hamiltonians
\begin{equation}\label{eqn:Hamiltonian}
  H(\theta,\phi)=\sin{\theta}\cos{\phi} F_1+\sin{\theta}\sin{\phi} F_2+\cos{\theta} F_3
\end{equation}
for each class (see Supplemental Material Sec. \RNum{2} A \cite{Supplementary}). Here $(\theta,\phi)$ defines the unit vector $\vec{h}=(\sin{\theta}\cos{\phi},\sin{\theta}$ $\sin{\phi},\cos{\theta})$. Then we measure the expectation values of the three observables with respect to these ground-states. Figure \ref{fig:exp_results} illustrates the experimental results for the exemplary JNRs of eight classes. The experimental results are in line with the theoretical predictions, as they are very closed to the surface of the theoretical predictions. The average similarity $S$ between experimentally measured probability distributions $\{p_j\}$ and the theoretical values $\{p_j^{th}\}$ is above $0.994$, with $S=(\sqrt{p_0p_0^{th}}+\sqrt{p_1p_1^{th}}+\sqrt{p_2p_2^{th}})^2$. The convex hulls of the experimental data also show the same geometrical features (the number of $s$ and $e$). Deviations between the observed data and the theoretical values are mainly attributed to systematic errors in the settings of experimental parameters (see Supplemental Material Sec. \RNum{2} E \cite{Supplementary}). Apart from the first class, the rest seven classes of the JNR are different from either a sphere or an ellipsoid, showing distinct properties of qutrit state space from those of qubits.
\begin{figure*}
\centering
\includegraphics[width=\linewidth]{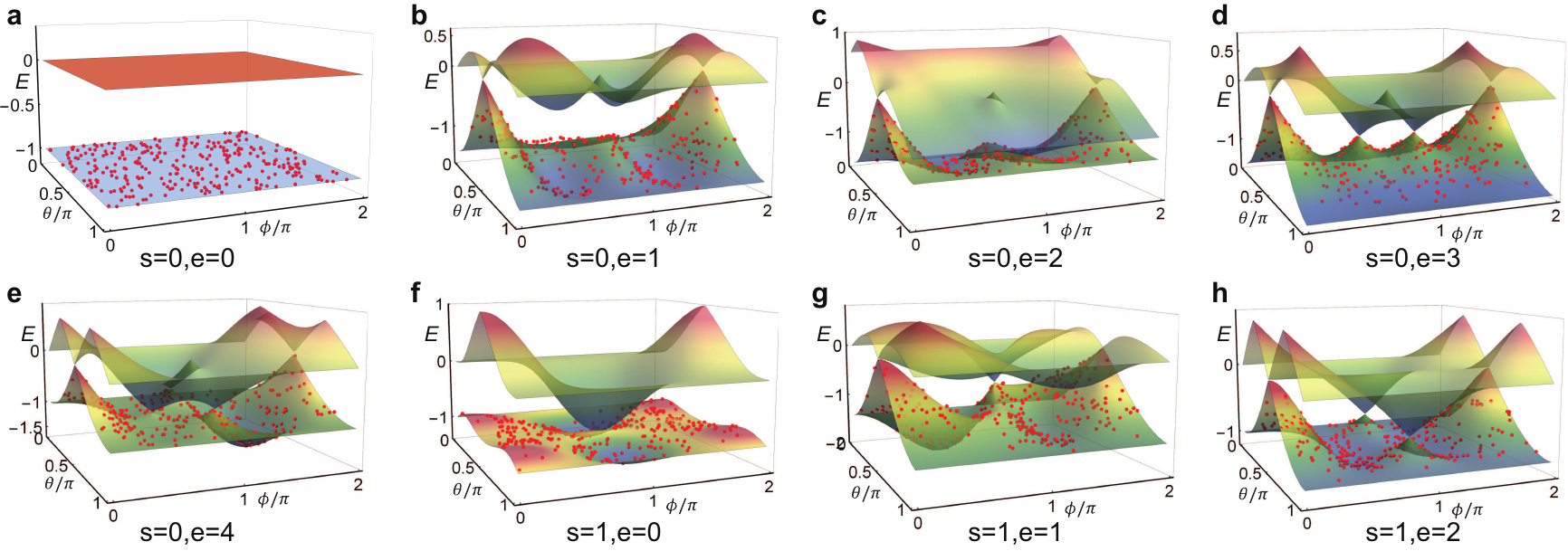}
\caption{\label{fig:band}Determining ground-state energy and degeneracy. (a)-(h) The band structures of the lowest two bands for the system Hamiltonians of the eight JNR classes. Colored surfaces represent the theoretical energy of the ground-state and the first excited state of the Hamiltonian $H(\theta,\phi)$ (Eq. (\ref{eqn:Hamiltonian})) for varied $\theta,\phi$. The experimental results (red dots) are computed by the experimental data of each class of JNR, as indirectly measured expectation values of $H(\theta,\phi)$. The number and features of degeneracy points shown in the band structure diagrams have a correspondence with the number of segments and ellipses on the surfaces of JNR.}
\end{figure*}

Following the complete observation of the geometric bodies, next we show how the geometry of JNRs in Fig. \ref{fig:exp_results} determines the ground-state energies and degeneracies of the system Hamiltonians. In Fig. \ref{fig:band}, by combining all the experimental results $(\langle F_1\rangle,\langle F_2\rangle,\langle F_3\rangle)$ of $\mathcal{F}$ with their corresponding unit vector $(\theta,\phi)$, we obtain the expectation value $E$ of the system Hamiltonian $H(\theta,\phi)$ and give the diagrams of $E$ (red dots) versus $\theta$ and $\phi$. The resulting energies are in line with theoretical prediction of ground-state energies (the lower colored surfaces) within experimental errors. In particular, there is a clear correspondence between the segments and ellipses in Fig. \ref{fig:exp_results} and the degeneracy points in Fig. \ref{fig:band}. For example, the first class (Fig. \ref{fig:band}(a)) corresponds to a gaped Hamiltonian and cannot see any flat portions from the surface of its JNR. As for the case $s=1,e=2$, there are three degeneracy points in the band structure diagram. Two are cone-shaped which are non-analytic in all directions and corresponds to the two ellipses on the JNR. The other is $\Lambda$-shaped which is non-analytic only in one direction and corresponds to the segment. These various band structure diagrams, which can be geometrically revealed by the surfaces of JNRs, also show distinct features of three-level quantum systems compared to two-level systems.
\par
For quantum matters at the zero temperature, the ground-states associated to a class of Hamiltonians are regarded as `quantum phases' of the matter. The degeneracy of ground-states, usually indicating a gap closing, is an important indicator of different quantum phases~\cite{PhysRevB.41.9377}, such as symmetry breaking, topological ordered and gapless phases. As demonstrated in our experiment, the flat surfaces (ellipses and segments in the case $d=n=3$) on an image of JNR indicate ground-state degeneracies of the Hamiltonian $H(\vec{h})$, thus represent different phases of the system. Therefore, the surface of JNR can be viewed as an intuitive geometric representation of quantum phases, on which different regions represent different phases.
\par
We experimentally identify the geometry of a three-level quantum system by observing the complete classification of joint numerical ranges (JNRs) of three observables. The results highlight the distinct difference between high-dimensional quantum systems and qubits. Furthermore, we elucidate the relation between the geometric characters of the JNR and ground-state degeneracies of the system Hamiltonians. Our work opens the avenue to experimentally explore fascinating phenomena of quantum systems via its state space projections. This geometric method is also applicable to many-body systems, when the set of observables are happen to be local Hamiltonians of the system, which may be easier to implement than the global ones~\cite{Kokail2019}. Besides, the concept of JNR has shown a wide range of potential applications. As convex sets, JNRs have been adopted in the $N$-representability problem~\cite{Erdahl1972,Zauner_2016} and the quantum marginal problem~\cite{Klyachko_2006} for visualizing the set of reduced density matrices. In the field of quantum information, it finds applications in the derivation of entanglement witnesses~\cite{PhysRevA.100.042326,PhysRevLett.110.060405} and uncertainty relations~\cite{Szyma_ski_2019,PhysRevLett.119.170404}, and provides theoretical foundations for the self-characterization of quantum devices~\cite{PhysRevLett.123.140405,PhysRevLett.124.040402}. We expect this versatile concept to promote further investigations in understanding the geometry of quantum systems, inferring intriguing phases and properties of quantum matters, as well as developing novel technologies in quantum information science.

\begin{acknowledgments}
The authors thank Y. Zhao, H. Zhang, and T. Lan for enlightening discussions. This work was supported by the National Key Research and Development Program of China (grant nos. 2017YFA0303703 and 2018YFA030602) and the National Natural Science Foundation of China (grant nos. 11690032, 91836303, 61490711, 11474159 and 11574145).
N.C. was supported by Natural Sciences and Engineering Research Council (NSERC) of Canada.
Research of Sze was supported by a HK RGC grant PolyU 15305719.
\end{acknowledgments}
%

\clearpage
\begin{widetext}
\section{Supplemental Material}

\section{Detailed Theoretical Information}
\subsection{Numerical range and its extensions}


Mathematically, Numerical Range (NR) $W(F)$ of a $d \times d$ matrix $F$ is $\left\{\bra{x}A\ket{x}, \ket{x}\in \mathbb{C}^d, \langle x|x\rangle=1\right\}$. Physicists are generally more interested in the case of $F$ being a Hermitian matrix, i.e. $F = F^\dagger$. It is straightforward to observe that $W(F)$ is the set of expectation values by measuring the observable $F$ with pure quantum states. The first important result of NR is called Toeplitz-Hausdorff theorem which states NRs are convex and compact\cite{Toeplitz1918,Hausdorff1919}.

A natural generalization of Numerical Range is that, instead of measuring one observable, one may measure a set of linearly independent operators $\mathcal{F} = \left\{F_1,\cdots,F_n\right\}$. This is so called Joint Numerical Range (JNR)
\begin{equation}
W(\mathcal{F}) = \left\{ (\bra{x}F_1\ket{x},\bra{x}F_2\ket{x},\cdots,\bra{x}F_n\ket{x}), \ket{x}\in \mathbb{C}^d, \langle x|x\rangle=1\right\}.
\end{equation}
The JNR of $\left\{F_1,F_2\right\}$ can be merged into NR since it equals to the NR of $A = F_1+\text{i}F_2$. 
Therefore, people usually consider $n \ge 3$ while studying JNR.
In contrast with Numerical Range, JNR is usually not convex~\cite{Au-Yeung197985,li2000convexity}.
In other words, the set of measurement outcomes that using only pure states to measure a set of observables may not be convex.
The next problem is how to characterize measurements according to mixed states.

In 1979, Au-Yeung and Poon~\cite{Au-Yeung197985} studied a generalized JNR
\begin{equation}\label{eq:JMR}
W^{(r)}(\mathcal{F}) = \left\{ \left(\sum_{i=1}^{r}\bra{x_i}F_1\ket{x_i},\sum_{i=1}^{r}\bra{x_i}F_2\ket{x_i},\cdots,\sum_{i=1}^{r}\bra{x_i}F_n\ket{x_i}\right) \bigg | \sum_{i=1}^{r}\langle x_i|x_i\rangle=1, \ket{x_i}\in \mathbb{C}^d \right\}.
\end{equation}
When $r=1$, it reduces to standard JNR where $r$ denotes the largest degree of quantum states used to measure the operator set $\mathcal{F}$.
Notice that $\ket{x_i}$ in~\cref{eq:JMR} is not normalized.
The probability $p_i$ of each pure state component is absorbed in $\ket{x_i}$.
Rewrite $\ket{x_i}$ as $\sqrt{p_i}\ket{x_i'}$, where $\langle x_i|x_i\rangle = 1$ and $\sum_i p_i =1$,
We can reformulate $W^{(r)}(\mathcal{F})$ as

\begin{equation}
W^{(r)}(\mathcal{F}) = \left\{ \left(\tr(F_1\rho),\tr(F_2\rho),\cdots,\tr(F_n\rho)\right)\bigg | \sum_{i=1}^{r}p_i|x_i'\rangle\langle x_i'|=\rho, \sum_{i=1}^{r}p_i =1, \langle x_i'|x_i'\rangle = 1, \ket{x_i'}\in \mathbb{C}^d, p_i\in [0,1]\right\}.
\end{equation}

In~\cite{Au-Yeung197985}, a lower bound for $W^{(r)}(\mathcal{F})$ to be convex has been provided.
For the scenario considered in this paper, which is qutrit systems ($d=3$) with three linearly independent observables ($n=3$), the measurement results of pure states (JNR of $\mathcal{F}$, $W(\mathcal{F})$) is convex.
Moreover, in qutrit systems, if we allow the use of mixed states up to degree $2$, then $W^{(r)}(\mathcal{F})$ is convex whenever $|\mathcal{F}| <8$.

Another concept introduced in \cite{BD} is the joint algebraic numerical range (JANR)

\begin{equation} 
L(\mathcal{F})= \left\{ \left(\tr(F_1\rho),\tr(F_2\rho),\cdots,\tr(F_n\rho)\right)\bigg | \rho \mbox{ is a quantum state}\right\}.
\end{equation}
Clearly, $L(\mathcal{F})=\text{Cov}(W(\mathcal{F}))$.
Since $W(\mathcal{F})$ is convex for
 the case discussed in this paper ($d=n=3$),  we do not distinguish the difference between the JANR and JNR in the following sections.

\subsection{The structure of joint algebraic numerical range}\label{Sec:JANR}

In this section, we discuss the structure of joint algebraic numerical range (or the convex hull of JNR).

Firstly, we show that the boundary of JANR of a linearly independent  set $\mathcal{F}$ of operators can be reached  by ground-states of Hamiltonians $H = \sum_{i=1}^n h_iF_i$, where $( h_1,\cdots, h_n) \in \mathbb{R}^n$. Firstly, it is obvious that the point $(\langle F_1\rangle,...,\langle F_n\rangle)$ with minimum value of $\langle F_1\rangle$ can be obtained by the ground-state of $F_1$, which is equivalent to making a tangent (hyper)plane of $L(\mathcal{F})$ with inward-pointing normal vector $(1,...,0)$. The resulting is the wanted extreme point. This tangent point also corresponds to the point obtained by the ground-state of $H((1,...,0))=F_1$. Therefore, In order to get other surface points, one only need to rotate the (hyper)plane with inward-pointing normal vector $(1,...,0)$ to other direction $\vec{h}$ and tangent to $L(\mathcal{F})$. This new tangent point can be obtained by the ground-state of a rotated operator $F'=\vec{h}\cdot \{F_1,...,F_n\}$, which is exactly the system Hamiltonian $H(\vec{h})$. And we call the tangent (hyper)plane with normal vector $\vec{h}$ as the supporting hyperplane, denoted as $\Pi$. Thus the ground-state energy $E_g$ can  be written as $E_g=\vec{h}\cdot(\langle F_1\rangle_{\psi_g},...,\langle F_n\rangle_{\psi_g}) $.

Notice that $H$ is fully parameterized by $( h_1,\cdots, h_n)$ in terms of the operator set $\mathcal{F}$.
We thus represent the Hamiltonian by the parameter vector $\vec h = (h_1,\cdots, h_n)$.
Observe that $\vec h =  (h_1,\cdots, h_n)$ has the same eigenspaces with $-\vec h$ and the eigenvalues differ from one anther with a constant $-1$. In other words, the ground-state of $\vec h$ becomes the highest excited states of $-\vec h$.
Since $\vec h$ and $-\vec h$ are all in $\mathbb{R}^n$, the image of ground-states of such class of $H$ ($H = \sum_{i=1}^n h_iF_i$) is the same as the image of highest excited states of $H$. The same discussion could be extended to the second excited states and the second highest excited states etc.
Therefore, one only need to study $\lceil \frac d2 \rceil$ levels of the eigenspaces and points generated by excited states are generally inside the JANR, apart from the highest excited states. For example, \cref{fig:nested} is the structure of the first three energy levels of a qudit system for which $d = 9$ and the set $\lb F_1, F_2\rb$ was random generated. The different energy levels show a nested structure, which is simply due to the order of eigen-energies. As for the mixed states, it is evident that points generated by them should inside the JANR.

In terms of the geometry of JNR, people are more interested in ground-states which form the surface.
If the ground-state of $H$ is not degenerate, there is only one point $(\val{F_1}_\psi, \val{F_2}_\psi, \cdots, \val{F_n}_\psi)$ on the surface of JANR corresponding to the ground-state.

Whereas, as for excited states, this is generally not the case. For one particular excited energy level, it forms a "fish" shape structure inside the JANR (see \cref{fig:nested}). Most of the points on the orbit are identifiable from each other, only a few points are overlapping at the "fish tail". That means eigen-states for the same excited level of different Hamiltonians may give the same measurement results of $\mathcal{F}$.

\begin{figure}
\includegraphics[width = 0.5\linewidth]{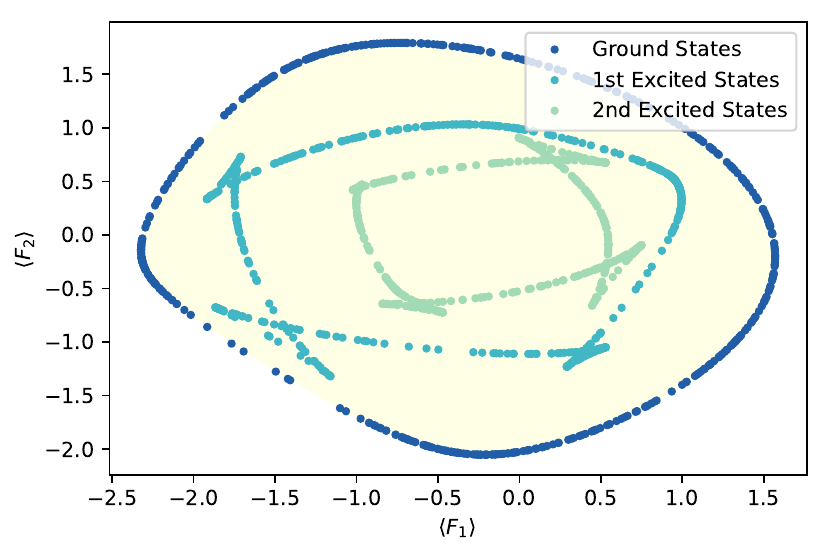}
\caption{The Joint numerical range of two $9\times9$ random generated hermitian operators $\lb F_1,F_2 \rb$.
($d = 9$ and $n=2$) The colored area is the JNR $W(\mathcal{F})$ of $\mathcal{F}=\lb F_1,F_2 \rb$. The blue dots are the measurements according to ground-states of $H = h_1F_1+h_2F_2$, which $h_1$ and $h_2$ are random generated. The cyan dots is formed by the first excited states of $H$. The green dots correspond to the second excited states.}
\label{fig:nested}
\end{figure}

\subsection{Ground-state degeneracy and the geometry of JNR surface}

The surface of joint numerical range $W(\mathcal{F})$ is determined by the ground-states of the class of Hamiltonians associated with $\mathcal{F}$.
We considered the geometry of JNRs of three $3\times3$ Hermitian matrices in this paper. Suppose  $\mathcal{F}=\left\{F_1,F_2, F_3 \right\}$ where $\left\{F_1,F_2,F_3, I_d\right\}$ is linearly independent.
The analysis of the influences of ground-state degeneracy on the surface of JNRs is given as follows:

Consider a supporting  plane $\Pi$ of $W(\mathcal{F})$ with inward-pointing normal vector  $\vec h = (h_1, h_2, h_3)$.  Suppose $H = \sum_{i=1}^3 h_iF_i$ has degenerate ground-states. Then the degeneracy is 2-dimensional. Choose two ground-states, $\ket{x_1}$ and $\ket{x_2}$, orthogonal to each other.
Let $X$ be the $3\times 2 $ matrix with columns $\ket{x_1}$ and $\ket{x_2}$. Let $\cS$ be the linear span of the set of $2\times 2 $ matrices $\{X^*F_iX:1\le i \le 3\} \cup \{I_2\}$.  Then $1\le $ dim $\cS\le 3$. We have\begin{enumerate}
\item If dim $\cS=1$, then $\Pi$ touches $W(\cF)$ at a point.

\item If dim $\cS=2$, then   $\Pi\cap  W(\cF)$ is a non-degenerate line segment.
\item If dim $\cS=3$, then   $\Pi\cap W(\cF)$ is a non-degenerate ellipse.
\end{enumerate}


\begin{figure}[t]
     \centering
     \begin{subfigure}{0.24\linewidth}
         \centering
         \includegraphics[width=\linewidth]{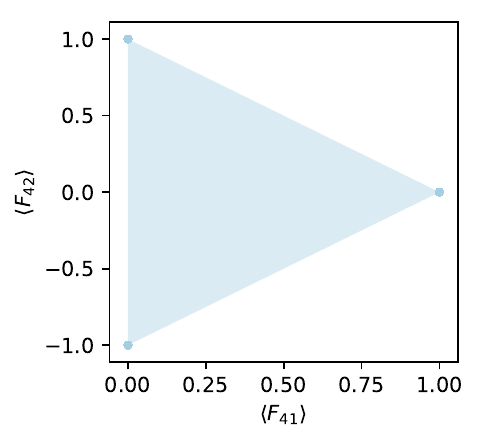}
         \caption{$W(F_{41},F_{42})$}
         \label{fig:AB}
     \end{subfigure}
     \begin{subfigure}{0.24\linewidth}
         \centering
         \includegraphics[width=\linewidth]{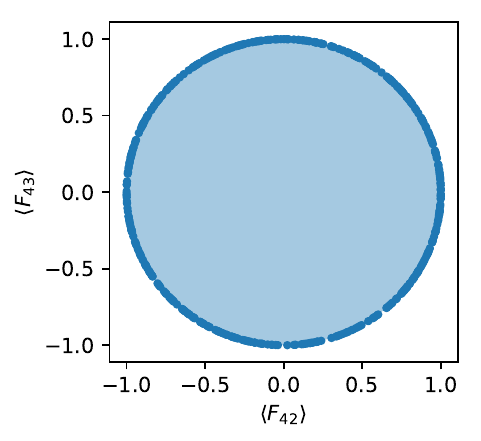}
         \caption{$W(F_{42},F_{43})$}
         \label{fig:BC}
     \end{subfigure}
     \begin{subfigure}{0.24\linewidth}
         \centering
         \includegraphics[width=\linewidth]{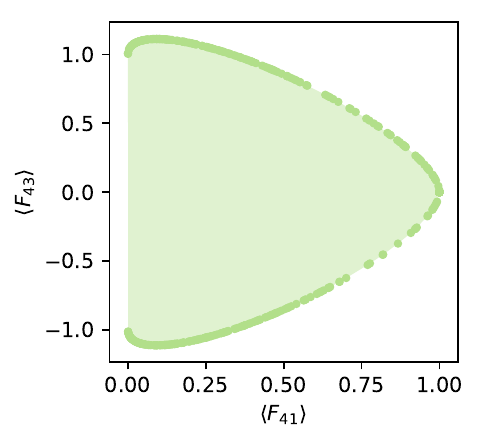}
         \caption{$W(F_{41},F_{43})$}
         \label{fig:AC}
     \end{subfigure}
     \begin{subfigure}{0.25\linewidth}
         \centering
         \includegraphics[width=0.8\linewidth]{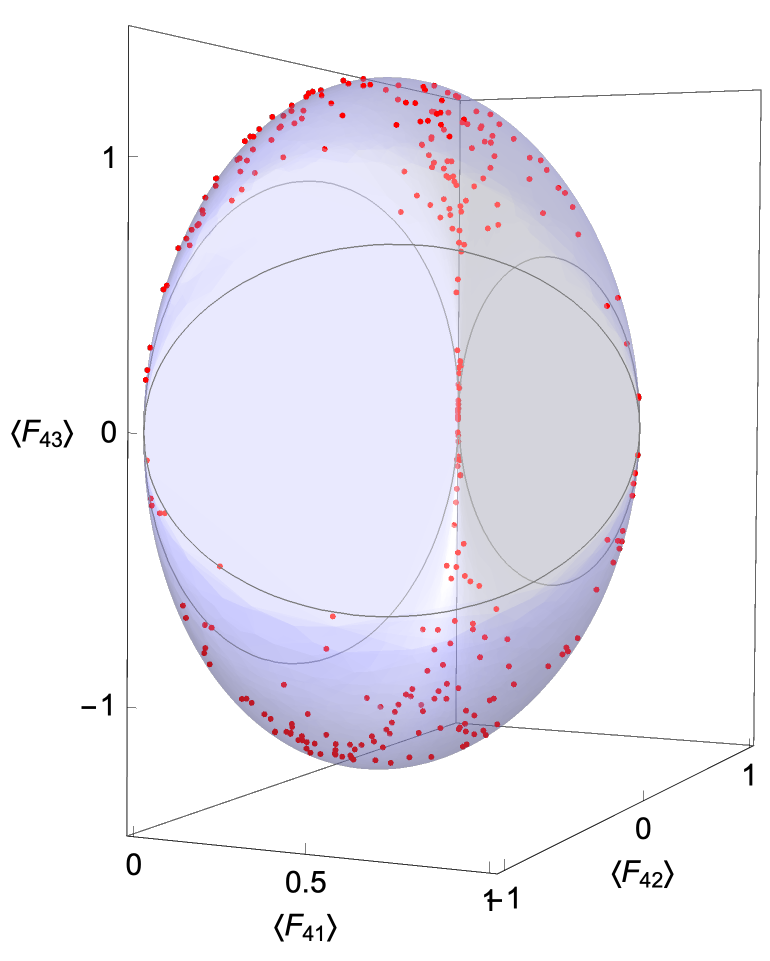}
         \caption{$W(\mathcal{F}_4)$}
         \label{fig:W4}
     \end{subfigure}
         \begin{subfigure}{0.24\linewidth}
         \centering
         \includegraphics[width=\linewidth]{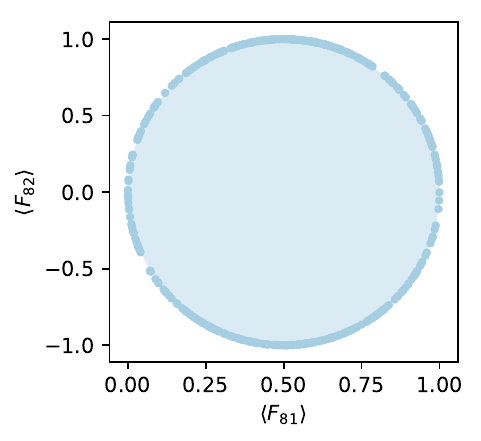}
         \caption{$W(F_{81},F_{82})$}
         \label{fig:AB8}
     \end{subfigure}
     \begin{subfigure}{0.24\linewidth}
         \centering
         \includegraphics[width=\linewidth]{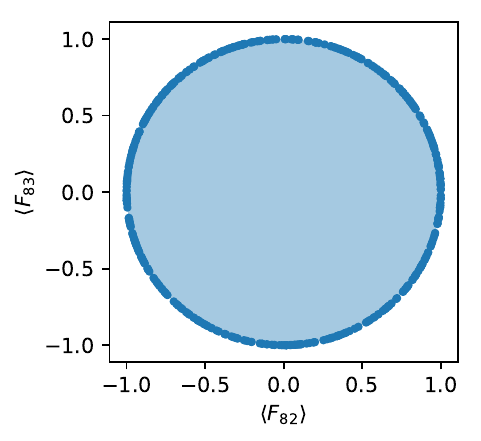}
         \caption{$W(F_{82},F_{83})$}
         \label{fig:BC8}
     \end{subfigure}
     \begin{subfigure}{0.24\linewidth}
         \centering
         \includegraphics[width=\linewidth]{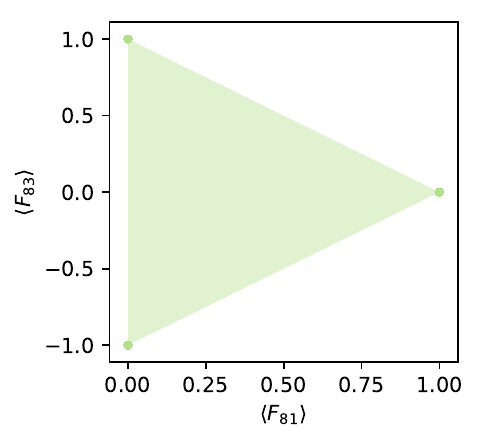}
         \caption{$W(F_{81},F_{83})$}
         \label{fig:AC8}
     \end{subfigure}
     \begin{subfigure}{0.24\linewidth}
         \centering
         \includegraphics[width=0.8\linewidth]{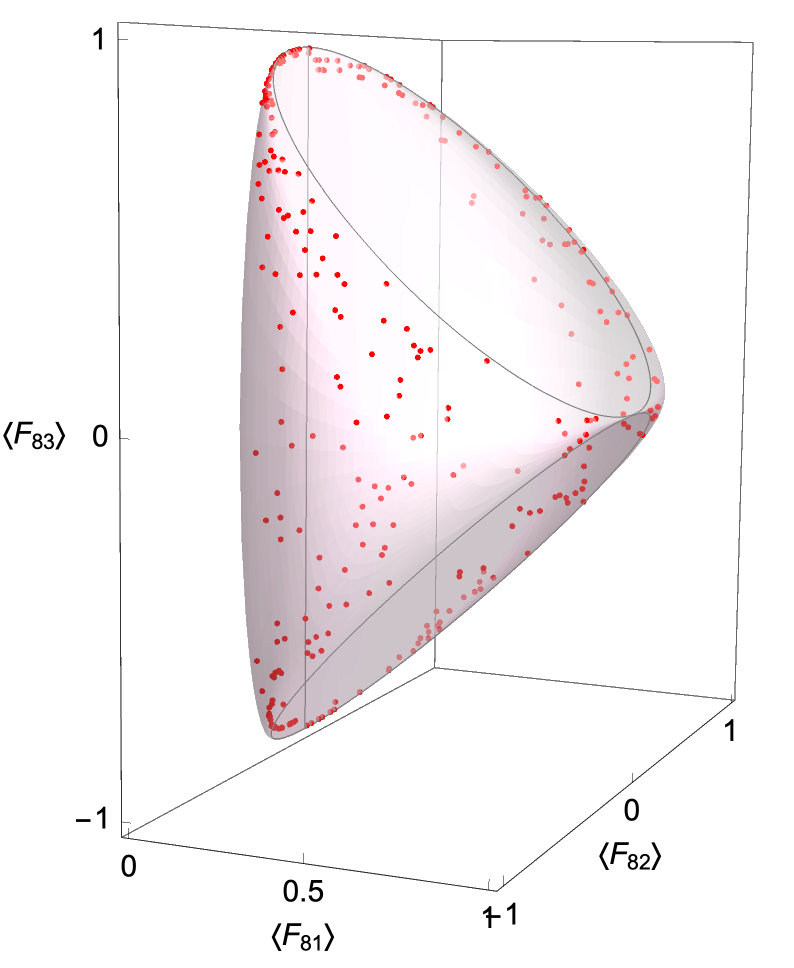}
         \caption{$W(\mathcal{F}_8)$}
         \label{fig:W8}
     \end{subfigure}
        \caption{The Joint Numerical Ranges of Class 4 and Class 8 and their projections to three coordinate planes. }
        \label{fig:C4C8}
\end{figure}

The examplary JNR of Class 4 ($e=3.s=0$) and Class 8 ($e=2,s=1$) used in this paper are appropriate examples for scenario 2 and scenario 3. \cref{fig:C4C8} depicts their JNRs and 2-D projections.
For Class 4, the set of $\mathcal{F}$ is (see also in ~\cref{tab:matrix})
\[
F_{41}=\(\begin{array}{ccc}0&0&0\\ 0&0&0\\ 0&0&1\end{array}\), F_{42}=\(\begin{array}{ccc}0&1&0\\ 1&0&0\\ 0&0&0\end{array}\), F_{43}=\(\begin{array}{ccc}0&i&1\\ -i&0&0\\ 1&0&0\end{array}\).
\]
The projection of $W(\mathcal{F})$ to the $x$-$y$ plane (\cref{fig:AB}, $W(F_{41},F_{42})$) is a triangle with inward-pointing normal vectors $(1,0)$, $(-1, -1)$ and $(-1,1)$ showing the degeneracy for $\vec h =(1,0,0), (-1,-1,0) $ and $(-1,1,0)$.
The corresponding $X$ is given by
\[
\(\begin{array}{ccc}1&0 \\ 0&1 \\ 0&0 \end{array}\), \frac{1}{\sqrt{2}}
\(\begin{array}{ccc}1&0 \\ 1&0 \\ 0&\sqrt{2}\end{array}\), \frac{1}{\sqrt{2}}
\(\begin{array}{ccc}1&0 \\ -1&0 \\ 0&\sqrt{2}\end{array}\)
\]
respectively. Direct calculation shows that dim $\cS=3 $ in all three degenerate points. Therefore, $W(\cF_4)$ has three ellipses on its surface.

The same discussion holds for $W(\mathcal{F}_8)$.
For Class 8, we have (see also in ~\cref{tab:matrix})

\[F_{81}=\(\begin{array}{ccc}0&0&0\\ 0&0&0\\ 0&0&1\end{array}\),
F_{82}=\(\begin{array}{ccc}0&0&1\\ 0&0&0\\ 1&0&0\end{array}\),
F_{83}=\(\begin{array}{ccc}0&1&0\\ 1&0&0\\ 0&0&0\end{array}\)
\]
The projection of $W(\mathcal{F})$ to the $x$-$z$ plane (\cref{fig:AC8}, $W(F_{81},F_{83})$) is a triangle with inward-pointing normal vectors $(1,0)$, $(-1, -1)$ and $(-1,1)$ showing the degeneracy for $\vec h =(1,0,0), (-1,0,-1) $ and $(-1,0,1)$. The corresponding $X$ is given by
\[
\(\begin{array}{ccc}1&0 \\ 0&1 \\ 0&0 \end{array}\),
\frac{1}{\sqrt{2}}
\(\begin{array}{ccc}1&0 \\ 1&0 \\ 0&\sqrt{2}\end{array}\),
\frac{1}{\sqrt{2}}
\(\begin{array}{ccc}1&0 \\ -1&0 \\ 0&\sqrt{2}\end{array}\)
\]
respectively. Direct calculation shows that dim $\cS=2 $ in the first degenerate point and dim $\cS=3 $ in last two degenerate points. Therefore, $W(\cF_8)$ has one line segment and two ellipses on its surface.

For scenario 1, we provide the operator set $\mathcal{A}$ in~\cref{tab:class1} as an example. That is, the ground-state degeneracy of $H = \sum_i h_iA_i$ appears as a point on the surface of JNR when $\vec h = (1,0,0)$ and it can be easily checked that dim $\cS=1 $ at this degenerate point. Apart from this point, there is no other degenerate point of this system Hamiltonian. So $\mathcal{A}$ belong to the first JNR class - there are no lines and ellipses on the surfaces of its JNR. However, the operator set that we measured in our experiment is not $\mathcal{A}$ but $\mathcal{B}$ in~\cref{tab:class1} (see also the first class in~\cref{tab:matrix}), and operator set $\mathcal{B}$ also belong to class 1 but with no degeneracy. The existence of such a special operator set $\mathcal{A}$ is an interesting phenomenon to study, but beyond the scope of this work.

\begin{table}
\begin{center}
\begin{tabular}{|c|c|c|c|}
\hline
$\mathcal{A}$ & $A_1 = \begin{pmatrix} 0 & 0 & 0 \\ 0 & 0 & 0 \\ 0 & 0 & 1\end{pmatrix}$ &
$A_2 = \begin{pmatrix} 0 & 0 & 1 \\ 0 & 0 & 0 \\ 1 & 0 & 0\end{pmatrix}$ &
$A_3 = \begin{pmatrix} 0 & 0 & 0 \\ 0 & 0 & 1 \\ 0 & 1 & 0\end{pmatrix}$\\ \hline
$\mathcal{B}$ & $B_1 = \begin{pmatrix} 0 & 0 & 1 \\ 0 & 0 & 0 \\ 1 & 0 & 0\end{pmatrix}$ &
$B_2 = \begin{pmatrix} 0 & 0 & \text{i} \\ 0 & 0 & 0 \\ -\text{i} & 0 & 0\end{pmatrix}$ &
$B_3 = \begin{pmatrix} 0 & 0 & 0 \\ 0 & 0 & 1 \\ 0 & 1 & 0\end{pmatrix}$\\
\hline
\end{tabular}
\end{center}
\caption{Two sets of hermitian matrices of Class 1.}
\label{tab:class1}
\end{table}




\section{Experimental Details}
\subsection{State preparation of photonic qutrit}
\textit{Single photon source.---}Light pulses with 830nm central wavelength from a ultrafast Ti:Sapphire laser (76MHz repetition rate; Coherent Mira-HP) are firstly frequency doubled in a beta barium borate ($\beta$-BBO) crystal. The second harmonics are then used to pump a 10 mm bulk potassium dihydrogen phosphate (KDP) crystal phase-matched for type-II collinear degenerate downconversion to produce photon pairs denoted as the signal and the idler. After the PBS, the idler mode is detected by a SPCM (Excelitas Technologies, SPCM-AQRH-FC with a detection efficiency about 49\%) as the trigger of the heralded single photon source, whereas the signal mode is directed towards the following set-up (shown in Fig. 2(c) in the main text).

\textit{Qutrit state preparation.---}We use a calcite beam displacer (BD), wave plates and two phase retarders (PRs) to prepare arbitrary pure qutrit states encoded in the polarization and spatial optical modes, depicted in the state preparation module in Fig. 2(c) in the main text. Single photons with horizontal polarization $\ket{H}$ firstly been scrambled in the superposition of horizontal and vertical polarization by the half-wave plate (HWP) $\rm{HWP}_A$ with setting angle $\theta_A$,
\begin{equation*}\label{eqn:psi1}
  \ket{\psi_1}=\cos{\theta_A}\ket{H}+\sin{\theta_A}\ket{V}.
\end{equation*}
Then followed by a PR, a relative phase between horizontal and vertical polarization is included and the state can be written as
\begin{equation*}\label{eqn:psi2}
  \ket{\psi_2}=e^{i\phi_1}\cos{\theta_A}\ket{H}+\sin{\theta_A}\ket{V}.
\end{equation*}
After BD1, the state is in the superposition of two spatial mode $\ket{s1}$ (top) and $\ket{s2}$ (lower),
\begin{equation*}\label{eqn:psi3}
  \ket{\psi_3}=e^{i\phi_1}\cos{\theta_A}\ket{H}\otimes\ket{s1}+\sin{\theta_A}\ket{V}\otimes\ket{s2}.
\end{equation*}
In the spatial mode $\ket{s1}$, a HWP with angle $45^\circ$ flip the polarization into $\ket{V}$, while in spatial mode $\ket{s2}$, $\rm{HWP}_B$ is set to $\theta_B$, therefore, the state after $\rm{HWP}_B$ is
\begin{equation*}\label{eqn:psi4}
  \ket{\psi_4}=e^{i\phi_1}\cos{\theta_A}\ket{V}\otimes\ket{s1}+\sin{\theta_A}(\sin{\theta_B}\ket{H}-\cos{\theta_B}\ket{V})\otimes\ket{s2}.
\end{equation*}
To generate arbitrary pure qutrit state, another relative phase still needed. Here we use the second PR simultaneously manipulating the two spatial mode, introducing another phase factor $\phi_2$ between horizontal and vertical polarization modes, thus in the end, when the photon pass through BD2, we have prepared an arbitrary qutrit state in the form of
\begin{equation*}\label{eqn:psi}
  \ket{\psi_5}=e^{i\phi_1}\cos{\theta_A}\ket{V}\otimes\ket{s1}+e^{i\phi_2}\sin{\theta_A}\sin{\theta_B}\ket{H}\otimes\ket{s1}-\sin\theta_A\cos{\theta_B}\ket{V}\otimes\ket{s2}.
\end{equation*}
By defining the three eigen-modes of the qutrit state as
\begin{equation*}\label{eqn:eigen_basis}
  \ket{0}\equiv\ket{H}\otimes\ket{s1},\ket{1}\equiv\ket{V}\otimes\ket{s1},\ket{2}\equiv\ket{V}\otimes\ket{s2},
\end{equation*}
the state of single photons go through the state preparation module can be written as
\begin{equation*}\label{eqn:psi}
  \ket{\psi}=e^{i\phi_2}\sin{\theta_A}\sin{\theta_B}\ket{0}+e^{i\phi_1}\cos{\theta_A}\ket{1}-\sin\theta_A\cos{\theta_B}\ket{2}.
\end{equation*}

For realistic implementation of the two PRs, we use different set-ups. The first PR is implemented by a liquid crystal phase retarder (Thorlabs, LCC1113-B) with its optical axis parallel to the horizontal polarization. Different phase $\phi_1$ can be realized by applying different voltage to this retarder. However, due to the tiny separating distance (4mm) of the two spatial modes and limited retardance uniformity of the liquid crystal phase retarder, the second PR is realized by a electronically-controlled HWP sandwiched by two quarter-wave plates (QWPs) (a QWP-HWP-QWP configuration) \cite{PhysRevLett.118.020403} with the HWP setting at $\frac{\phi_2}{4}$ an  the two QWPs setting at $45^\circ$, which equivalently performing the unitary operation below by using the Jones matrix notation \cite{PhysRevA.64.012303} of wave plates,
\begin{equation*}\label{eqn:U_QHQ}
  U_{QHQ}=e^{i(\pi-\phi)}\ket{H}\bra{H}+\ket{V}\bra{V}.
\end{equation*}
By electrically setting different angels to E-HWP, different $\phi_2$ can also be achieved. All the components in state preparation module are electronically-controlled or fixed, which ensure high repeatability in state preparation stage when measuring different observables.

\textit{Sampling surface states.---}Given a triple of Hermitian observables $\mathcal{F}=(F_1,F_2,F_3)$, the surface points of $L(\mathcal{F})$ can be derived by the ground-state of the system Hamiltonian
\begin{equation*}\label{eqn:supporting Hamiltonian}
  H=\sin{\theta}\cos{\phi} F_1+\sin{\theta}\sin{\phi} F_2+\cos{\theta} F_3,
\end{equation*}
where $(\theta,\phi)$ defines the orientation of the supporting hyperplane. In our experiment, we randomly sample $\theta\in[0,\pi]$ and $\phi\in[0,2\pi]$ to generate different $H$, and then calculate their ground-states as the target states we are going to prepare with the above mentioned set-ups. Totally, we sampled 300 this kind of surface states for each class of $\mathcal{F}$ and measured the expectation values of $F_1,F_2$ and $F_3$ to obtain the data points of $L(\mathcal{F})$.

\subsection{Measurement of joint numerical ranges}
\par
For the experimental observation of the classification of qutrit JNR, we experimentally measured the expectation values of 8 classes of triple Hermitian observables $\mathcal{F}=(F_1,F_2,F_3)$. For each classes, 300 surface points were measured by sampling 300 surface states. The 8 classes of exemplary Hermitian observables that we measured are shown in \cref{tab:matrix}.
\begin{table}[h!]
  \centering    \caption{\label{tab:matrix} 8 classes of $\mathcal{F}$ measured in our experiment. ($\bar{1}$ denote $-1$ and $\bar{i}$ denote $-i$)}
    \begin{tabular}{c|c|c<{\centering}}
      \hline
      Class & Feature  & $F$ \\
      \hline
      1 & $s=0,e=0$ & $F_{11}=\begin{pmatrix} 0 & 0 & 1 \\ 0 & 0 & 0 \\ 1 & 0 & 0\end{pmatrix},
      F_{12}=\begin{pmatrix} 0 & 0 & i \\ 0 & 0 & 0 \\ \bar{i} & 0 & 0\end{pmatrix},
      F_{13}=\begin{pmatrix} 0 & 0 & 0 \\ 0 & 0 & 1 \\ 0 & 1 & 0\end{pmatrix}$ \\
      \hline
      2 & $s=0,e=1$ & $F_{21}=\begin{pmatrix} 0 & 0 & 0 \\ 0 & 0 & 0 \\ 0 & 0 & 1\end{pmatrix},
      F_{22}=\begin{pmatrix} 0 & 1 & 0 \\ 1 & 0 & 1 \\ 0 & 1 & 0\end{pmatrix},
      F_{23}=\begin{pmatrix} 0 & i & 1 \\ \bar{i} & 0 & 0 \\ 1 & 0 & 0\end{pmatrix}$ \\
      \hline
      3 & $s=0,e=2$ & $F_{31}=\begin{pmatrix} 0 & 0 & 0 \\ 0 & 0 & 0 \\ 0 & 0 & 1\end{pmatrix},
      F_{32}=\begin{pmatrix} 0 & 1 & \bar{1} \\ 1 & 0 & 0 \\ \bar{1} & 0 & 0\end{pmatrix},
      F_{33}=\begin{pmatrix} 1 & 0 & 0 \\ 0 & \bar{1} & 1 \\ 0 & 1 & 0\end{pmatrix}$ \\
      \hline
      4 & $s=0,e=3$ & $F_{41}=\begin{pmatrix} 0 & 0 & 0 \\ 0 & 0 & 0 \\ 0 & 0 & 1\end{pmatrix},
      F_{42}=\begin{pmatrix} 0 & 1 & 0 \\ 1 & 0 & 0 \\ 0 & 0 & 0\end{pmatrix},
      F_{43}=\begin{pmatrix} 0 & i & 1 \\ \bar{i} & 0 & 0 \\ 1 & 0 & 0\end{pmatrix}$ \\
      \hline
      5 & $s=0,e=4$ & $F_{51}=\begin{pmatrix} 0 & 0 & 0 \\ 0 & 0 & 0 \\ 0 & 0 & 1\end{pmatrix},
      F_{52}=\begin{pmatrix} 1 & 0 & 1 \\ 0 & \bar{1} & 0 \\ 1 & 0 & 0\end{pmatrix},
      F_{53}=\begin{pmatrix} 0 & 1 & 0 \\ 1 & 0 & 0 \\ 0 & 0 & 0\end{pmatrix}$ \\
      \hline
      6 & $s=1,e=0$ & $F_{61}=\begin{pmatrix} 0 & i & 0 \\ \bar{i} & 0 & 0 \\ 0 & 0 & 1\end{pmatrix},
      F_{62}=\begin{pmatrix} 0 & 1 & 0 \\ 1 & 0 & 0 \\ 0 & 0 & 0\end{pmatrix},
      F_{63}=\begin{pmatrix} 0 & 0 & 1 \\ 0 & 0 & 0 \\ 1 & 0 & 0\end{pmatrix}$ \\
      \hline
      7 & $s=1,e=1$ & $F_{71}=\begin{pmatrix} 0 & 0 & 0 \\ 0 & 0 & 0 \\ 0 & 0 & 1\end{pmatrix},
      F_{72}=\begin{pmatrix} 0 & 0 & \bar{1} \\ 0 & 0 & 1 \\ \bar{1} & 1 & 0\end{pmatrix},
      F_{73}=\begin{pmatrix} 0 & 1 & 0 \\ 1 & 0 & 1 \\ 0 & 1 & 0\end{pmatrix}$ \\
      \hline
      8 & $s=1,e=2$ & $F_{81}=\begin{pmatrix} 0 & 0 & 0 \\ 0 & 0 & 0 \\ 0 & 0 & 1\end{pmatrix},
      F_{82}=\begin{pmatrix} 0 & 0 & 1 \\ 0 & 0 & 0 \\ 1 & 0 & 0\end{pmatrix},
      F_{83}=\begin{pmatrix} 0 & 1 & 0 \\ 1 & 0 & 0 \\ 0 & 0 & 0\end{pmatrix}$ \\
      \hline
    \end{tabular}
\end{table}

\begin{table}[h!]
  \centering    \caption{\label{tab:setting_angles} Wave plate setting angles for different non-diagonal observables. }
    \begin{tabular}{c|c|c|c|c<{\centering}}
      \hline
      Observable & HWP1  & QWP & HWP2 & HWP3 \\
      \hline
      $F_{11},F_{63},F_{82}$ & 90 & * & 90 & 112.5 \\
      \hline
      $F_{12}$ & 90 & 90 (before HWP1) & 90 & 112.5 \\
      \hline
      $F_{13}$ & 45 & * & 90 & 67.5 \\
      \hline
      $F_{22},F_{73}$ & 45 & * & 67.5 & 112.5 \\
      \hline
      $F_{23},F_{43}$ & 90 & 0 (after HWP1) & -67.5 & -67.5 \\
      \hline
      $F_{32}$ & 90 & * & 112.5 & 67.5 \\
      \hline
      $F_{33}$ & 90 & * & 60.86 & 45 \\
      \hline
      $F_{42},F_{53},F_{62},F_{83}$ & 67.5 & * & 45 & 90 \\
      \hline
      $F_{52}$ & 45 & * & 60.86 & 90 \\
      \hline
      $F_{61}$ & -67.5 & 0 (before HWP1) & 90 & 90 \\
      \hline
      $F_{72}$ & 67.5 & * & 90 & 67.5 \\
      \hline
 \end{tabular}
\end{table}

To measure the expectation value $\langle F_i\rangle$ of a Hermitian observable $F_i$ with respect to an arbitrary state $\ket{\psi}$, we first rewrite $\ket{\psi}$ in the eigen-basis of $F_i$,
\begin{equation*}\label{eqn:psi}
  \ket{\psi}=\sqrt{p_0}\ket{\lambda^{(i)}_0}+\sqrt{p_1}\ket{\lambda^{(i)}_1}+\sqrt{p_2}\ket{\lambda^{(i)}_2},
\end{equation*}
where $\ket{\lambda^{(i)}_j}(j=0,1,2)$ is the corresponding eigen-vector with eigen-value $\lambda^{(i)}_j$ of $F_i$ and $p_j(j=0,1,2)$ is the probability $\ket{\psi}$ been projected into the eigen-mode $\ket{\lambda^{(i)}_j}$. In order to measure the probability distribution $\{p_j\}$, we can apply a unitary transformation
\begin{equation*}\label{eqn:unitary}
  U_{F_i}=\ket{0}\bra{\lambda^{(i)}_0}+\ket{1}\bra{\lambda^{(i)}_1}+\ket{2}\bra{\lambda^{(i)}_2}
\end{equation*}
to $\ket{\psi}$, which transforms any state from the eigen-basis of $F_i$ into computational or experimental basis. Thus $\{p_j\}$ can be directly read out from the measurement statistics when projecting state $U_F\ket{\psi}$ into the three experimental basis. By defining detecting events in the experimental basis $\ket{j}(j=0,1,2)$ as measuring the outcome $\lambda_j^{(i)}$, then $\langle F_i\rangle$ can be derived by using $\langle F_i\rangle=\sum_j p_j^{(i)}\lambda_j^{(i)}$. To realize this unitary operation $U_{F_i}$, we implemented a three stage interferometer formed by BDs and wave plates, as shown in Fig 2c in the main text. In each stage, BDs and HWPs permutate two of the qutrit eigen-modes into the same spatial mode with different polarization, then the two modes were interfered by HWP and QWP, which equivalently performing a $2\times2$ unitary on the two modes and leaving the third mode unchanged. It has been shown that any $3\times3$ unitary operation $U$ can be written as $U=U_3U_2U_1$, where $U_1,U_2,U_3$ are of the form
\begin{equation*}\label{eqn:U1U2U3}
  U_1=\begin{pmatrix} m_1 & n_1 & 0 \\ p_1 & q_2 & 0 \\ 0 & 0 & 1\end{pmatrix},
  U_2=\begin{pmatrix} m_2 & 0 & n_2 \\ 0 & 1 & 0 \\ p_2 & 0 & q_2\end{pmatrix},
  U_3=\begin{pmatrix} 1 & 0 & 0 \\ 0 & m_3 & n_3 \\ 0 & p_3 & q_3\end{pmatrix},
\end{equation*}
 and $m_k,n_k,p_k,q_k(k=1,2,3)$ form a $2\times2$ unitary block. Therefore, by using the QWP-HWP-QWP configuration \cite{PhysRevLett.118.020403}, arbitrary $2\times2$ unitary block in $U_k$ can be realized in each interference stage, and in principle, this three stage interferometer can realize any $3\times3$ unitary $U$. In our experiments, the three stage interferometer from left to right perform unitary operations in the form of $U_1,U_2$ and $U_3$ sequentially. For most of the observables in \cref{tab:matrix}, only HWP is needed for the realization of $U_{F_i}$, and for some observables, QWP is needed, the wave plate setting angles of all the non-diagonal observables listed in \cref{tab:matrix} are show in \cref{tab:setting_angles}. As for diagonal observables, $U_{F_i}$ become the identity operator, and all the wave plates set to zero.

\subsection{Experimental observations of joint numerical ranges in the case $d=3,n=2$}
\begin{figure}[h!]
     \centering
     \begin{subfigure}{0.24\linewidth}
         \centering
         \includegraphics[width=\linewidth]{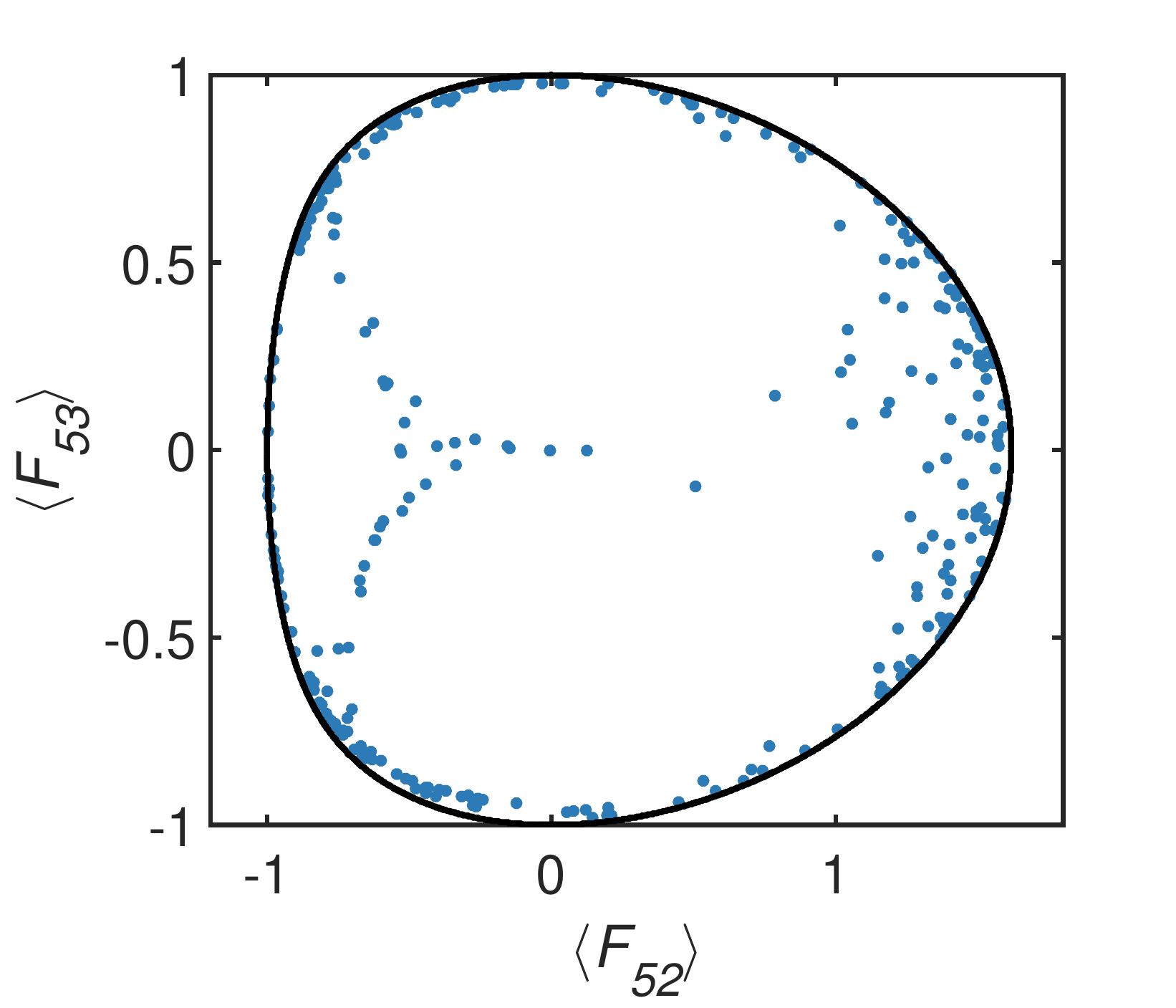}
         \label{fig:class1}
     \end{subfigure}
     \begin{subfigure}{0.24\linewidth}
         \centering
         \includegraphics[width=\linewidth]{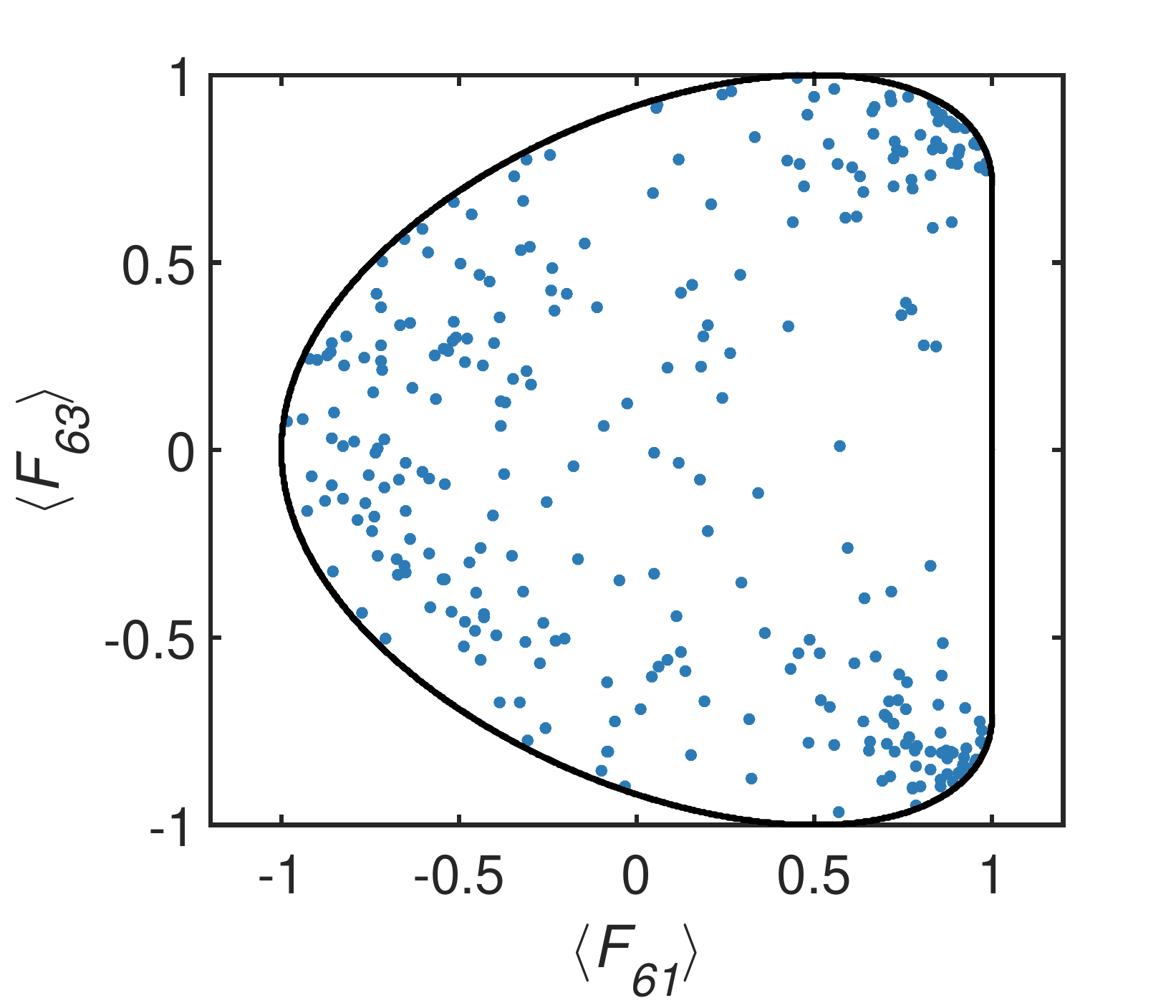}
         \label{fig:class2}
     \end{subfigure}
     \begin{subfigure}{0.24\linewidth}
         \centering
         \includegraphics[width=\linewidth]{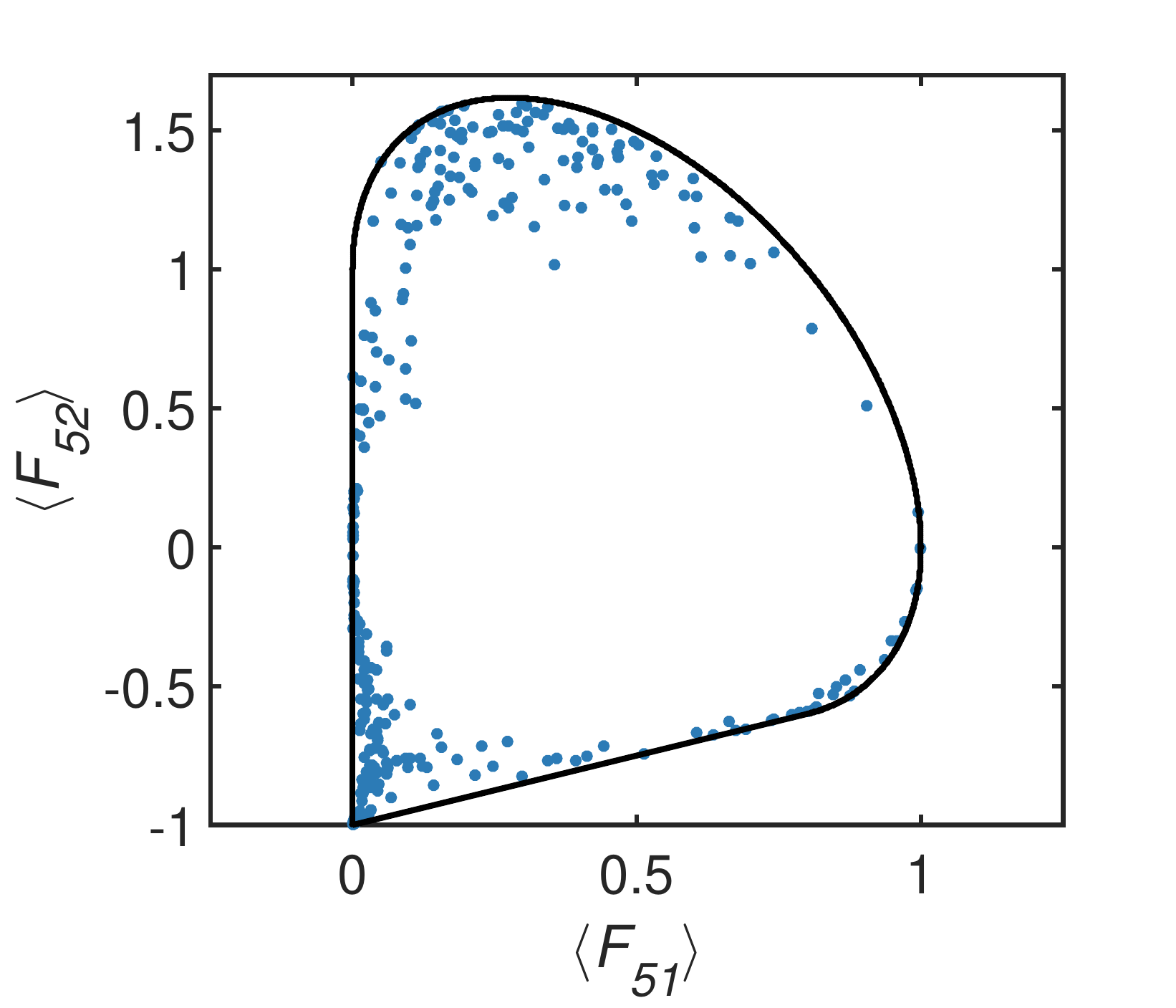}
         \label{fig:class3}
     \end{subfigure}
     \begin{subfigure}{0.24\linewidth}
         \centering
         \includegraphics[width=\linewidth]{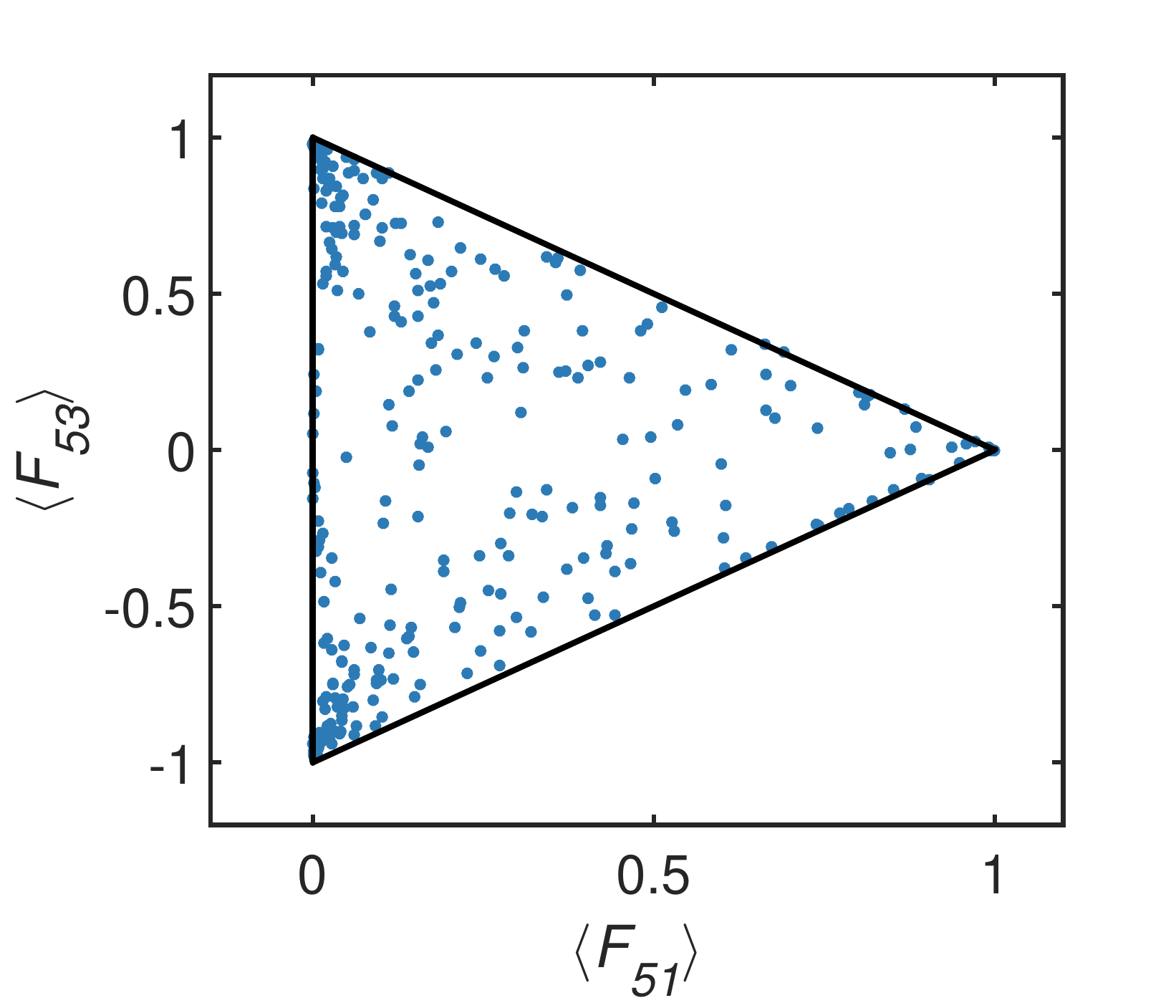}
         \label{fig:class4}
     \end{subfigure}
        \caption{Experimentally observed four classes of JNR of two $3\times3$ observables.  Theoretical boundary (two-dimensional surface) curves of these two dimensional plane sets are plotted with black lines while blue points (300 for each class) represent the experiment results. }
        \label{fig:JNR_2d}
\end{figure}

By projecting the three dimensional JNR in the main text into two dimensional coordinate plane, our experimental results also show an complete observation of the four classes of $L(\mathcal{F})$ in the case $d=3,n=2$~\cite{Keeler1997}. Four exemplary results are shown in \cref{fig:JNR_2d}. From left to right, the four classes are: an oval (the convex hull of a sextic curve), the convex hull of a quartic curve with a flat portion on the boundary, the convex hull of an ellipse and a point outside the ellipse, a triangle. As the boundary states of $L(\mathcal{F})$ with $d=3,n=2$ no longer being the surface states in the case $n=3$, most of the experimental points are inside the range, but still show well agreements with the theoretically predicted ranges.

\subsection{Two classes of unitarily reducible JNRs in the case $d=n=3$}
As mentioned in the main text, in the case $d=n=3$, there are ten possible categories of JNRs according to the number of ellipse $e$ and segment $s$. The experimental results of the eight unitarily irreducible classes are shown in the main text. The another two classes with $s=\infty,e=1$ and $s=\infty,e=0$, which correspond to linearly dependent set of $\mathcal{F}$, can be obtained by lower dimensional JNRs. For example, the JNR with $s=\infty,e=1$ can be generated by the JNR of the following two matrices
 \begin{equation*}
  X=\begin{pmatrix} 0 & 1 & 0 \\ 1 & 0 & 0 \\ 0 & 0 & 0\end{pmatrix},
  Y=\begin{pmatrix} 0 & -i & 0 \\ i & 0 & 0 \\ 0 & 0 & 0\end{pmatrix},
\end{equation*}
which is a circle belong the first class in \cref{fig:JNR_2d} and have the exact shape of the two-dimensional projection of a Bloch sphere on the $x-y$ plane. By simply adding a $z$ component that orthogonal to the nonzero two-dimensional subspace of $X$ and $Y$
 \begin{equation*}
  Z=\begin{pmatrix} 0 & 0 & 0 \\ 0 & 0 & 0 \\ 0 & 0 & 1\end{pmatrix},
\end{equation*}
it is not hard to imagine that the resulting JNR is a cone, with one ellipse and infinite number of segments. For each $\langle Z\rangle$ value, the cross section of the JNR is a circle and when $\langle Z\rangle$ get larger, the radius of the circle get smaller.


\subsection{Experimental error analysis}
As mentioned in the main text, deviations between the observed data and the theoretical values are mainly attributed to two kinds of systematic errors in the settings of experimental parameters. The first is the imperfection of the two PRs. The QWP-HWP-QWP configuration involvs three wave plates which suffering misalignments of the optics axis (typically $\sim0.1$ degree), retardation errors (typically $\sim\lambda/300$ where $\lambda=830$nm) and inaccuracies of setting angles (typically $\sim0.2$ degree) while the liquid crystal phase retarder was pre-calibrated by a co-linear interferometer formed by four wave plates which may transfer the experimental errors. Both cause inaccuracy in manipulating relative phase between horizontal and vertical polarizations and also cause inaccuracy in calibrating the three stage interferometers in the measurement part. The second kind of systematic errors come from the slowly drift and slight vibrating of the interferometers during the measurement progress, which cause a decreasing of interference visibility. Overall, the interference visibility during the whole measuring progress is above $98.7\%$ for all classes. The average similarity $S$ of experimentally measured probability distributions of all the operators are shown in \cref{tab:similarity}. All the average similarities are above 0.994, which indicates well overall performances of the experimental settings and high similarities between the experimental datas and theoretical distributions.

\begin{table}[h!]
  \centering    \caption{\label{tab:similarity} Average similarity $S$ of experimentally measured probability distributions. }
    \setlength{\tabcolsep}{4mm}
    \begin{tabular}{c|c|c|c|c|c|c|c|c<{\centering}}
      \hline
      Class & 1  & 2 & 3 & 4 & 5 & 6 & 7 & 8 \\
      \hline
      $S(F_1)$ & 0.9997 & 0.9998 & 0.9998 & 0.9997 & 0.9999 & 0.9979 & 0.9999 & 0.9997 \\
      \hline
      $S(F_2)$ & 0.9990 & 0.9985 & 0.9988 & 0.9986 & 0.9992 & 0.9986 & 0.9964 & 0.9948 \\
      \hline
      $S(F_2)$ & 0.9998 & 0.9992 & 0.9998 & 0.9991 & 0.9988 & 0.9996 & 0.9977 & 0.9955 \\
      \hline
 \end{tabular}
\end{table}

\end{widetext}

\end{document}